\documentclass[twocolumn,amsmath,amssymb,proc,secnumarabic,prl]{revtex4}

\usepackage{graphicx,cancel}
\usepackage{ulem}

\countdef\preprint=0 \countdef\tube=12
\ifnum\tube<10

\else

\fi

\def\e0{$E_0$}

\def\sqd{\sqrt{3}}

\begin{document}

\setlength{\mathindent}{0.2cm}

\title{Tunable hybridization of electronic states of graphene and a metal surface}

\author{Alexander~Gr\"{u}neis$^{1}$ and Denis Vyalikh$^{2}$}

\affiliation{$^{1}$IFW Dresden, P.O. Box 270116, D-01171 Dresden,
Germany}
\affiliation{$^{2}$ Institut f\"ur Festk\"orperphysik, TU
Dresden, Mommsenstrasse 13, D-01069 Dresden, Germany}

\date{\today}
\begin{abstract}
We present an approach to monitor and control the strength of the
hybridization between electronic states of graphene and metal
surfaces. Inspecting the distribution of the $\pi$ band in a
high-quality graphene layer synthesized on Ni(111) by
angle-resolved photoemission, we observe a new "kink" feature
which indicates a strong hybridization between $\pi$ and
\textit{d} states of graphene and nickel, respectively. Upon
deposition and gradual intercalation of potassium atoms into the
graphene/Ni(111) interface, the "kink" feature becomes less
pronounced pointing at potassium mediated attenuation of the
interaction between the graphene and the substrate.
\end{abstract}
\maketitle

Since the discovery of two--dimensional meta--stable graphene
sheets~\cite{geim07-review} much research has been devoted to
explore its electronic properties because it is most suitable for
nanoelectronic devices with high electronic mobilities of
(15000~$\rm cm^2V^{-1}s^{-1}$~\cite{novoselov06-graphite} at room
temperature) and can be lithographically patterned
successfully~\cite{berger07-graphene}. The electronic structure of
isolated graphene in the vicinity of the Fermi level (E$_F$) is
that of a zero–-gap semiconductor, where the bare bands are linear
and the density of states at E$_F$ equals zero. These exotic
properties give rise to a number of fascinating effects such as
quasi relativistic Klein tunneling~\cite{geim06-kleinparadox}, an
anomalous quantum Hall effect~\cite{novoselov06-graphite} and a
node in the optical absorption close to E$_F$~\cite{n950}. Strong
renormalizations of the quasi--particle bands due to
electron--electron~\cite{alex06-correlation},
electron--phonon~\cite{takahashi07-prl} and
electron--plasmon~\cite{rotenberg06-graphite} interactions were
observed by angle--resolved photoemission (ARPES) in pristine
graphite and graphene on SiC. ARPES spectra of graphite single
crystals are not affected by substrate (sample holder)
interactions since they exceed $\rm\sim 100 \mu m$ in thickness.
With the graphene layer, however, the SiC substrate interacts and
transfers electrons to the $\pi^*$ band of graphene. This charge
transfer occurs according to a rigid band shift model since there
is no hybridization of the Si and the C orbitals.

The situation is changed dramatically in the case of
graphene/metal systems. Strong hybridization of electronic states
and charge transfer between graphene and the metal substrate
modify the band structure of graphene. In this respect
intercalation of foreign atoms into those systems is a key point,
which can help to understand and separate these two competing
phenomena. It has already been demonstrated that graphene grown
epitaxially on Ni(111)~\cite{shelton74-graphene,oshima94-graphene}
has an electronic structure quite different to pristine graphene
for two reasons. First, the Ni $3d_{z^2-r^2}$ and C $2p_z$
orbitals are elongated along the direction perpendicular to the
graphene sheet and have a large overlap. Second, the interplane
distance between the graphene layer and the Ni(111) surface is
significantly smaller than in bulk graphite. Low energy electron
diffraction (LEED) measurements carried out on graphene/Ni(111)
system suggest that one atom of the graphene unit cell is located
on above the topmost Ni layer and the other carbon atom is located
on top of hollow sites~\cite{oshima97-grapheneposition}. This
structural assignment is consistent with the observation of a gap
opening at the $K$ point (or Dirac point) at the Brillouin zone
(BZ) corner~\cite{oshima94-graphene} which indicates that the two
atoms in the unit cell are situated at nonequivalent positions on
the Ni(111) surface. Also, it was found experimentally the K atoms
deposited onto the graphene surface penetrate to the
graphene/Ni(111) interface already at room
temperature~\cite{oshima94-graphene}. This approach offers an
appealing possibility to control the degree of hybridization
between Ni $3d_{z^2-r^2}$ and C $2p_z$ orbitals which is quite
important and strongly motivated by requirements of precise
control and functionalization of graphene based nanostructures.
Although a few works~\cite{oshima94-graphene,molo01-graphene} have
been conducted with the aim to study modifications of the
electronic structure of graphene on metallic substrates, the
detailed description of the hybridization phenomena in
graphene/metal interfaces is still lacking. Exploiting this
effect, however allows to explore charge transfer in the crossover
regime between hybridization and the simple rigid band model.

In this contribution we present a comprehensive and systematic
ARPES study of the evolution of the electronic band structure of
graphene layer on Ni(111) under gradual penetration of K atoms to
the graphene/Ni(111) interface. We show that due to strong
hybridization graphene and Nickel states the $\pi$ band is pushed
downwards by the Ni $3d$ bands, revealing new "kink" feature.
However, upon gradual deposition of K atoms the "kink"-like
structure becomes less pronounced indicating K-mediated
attenuation of the hybridization between the graphene and the
metallic substrate. In order to describe the changes in the
electronic structure quantitatively, we perform calculations on
the tight--binding (TB) level. The experiments were performed
using a photoemission spectrometer equipped with a Scienta SES-200
hemispherical electron-energy analyzer, a high-flux He-resonance
lamp (Gammadata VUV-5010) in combination with a grating
monochromator and an X-ray source. All valence PE spectra were
acquired at a photon energy of h$\nu$=40.8 eV (He II$\alpha$) with
an angular resolution of 0.3$^{\circ}$ and a total-system energy
resolution of 50 meV. The samples were measured at room
temperature. X--ray photoelectron spectroscopy (XPS) was applied
in order to estimate the K $2p$/C $1s$ intensity ratio (K/C) as
well as the quality of the samples. The XPS spectra were obtained
at a photon energy of 1486.6 eV (Al K$\alpha$). Cleanliness and
high crystalline quality of the prepared structures were cross
checked by LEED.

\begin{figure}[t]
\begin{tabular}{c}
\includegraphics[width=9cm]{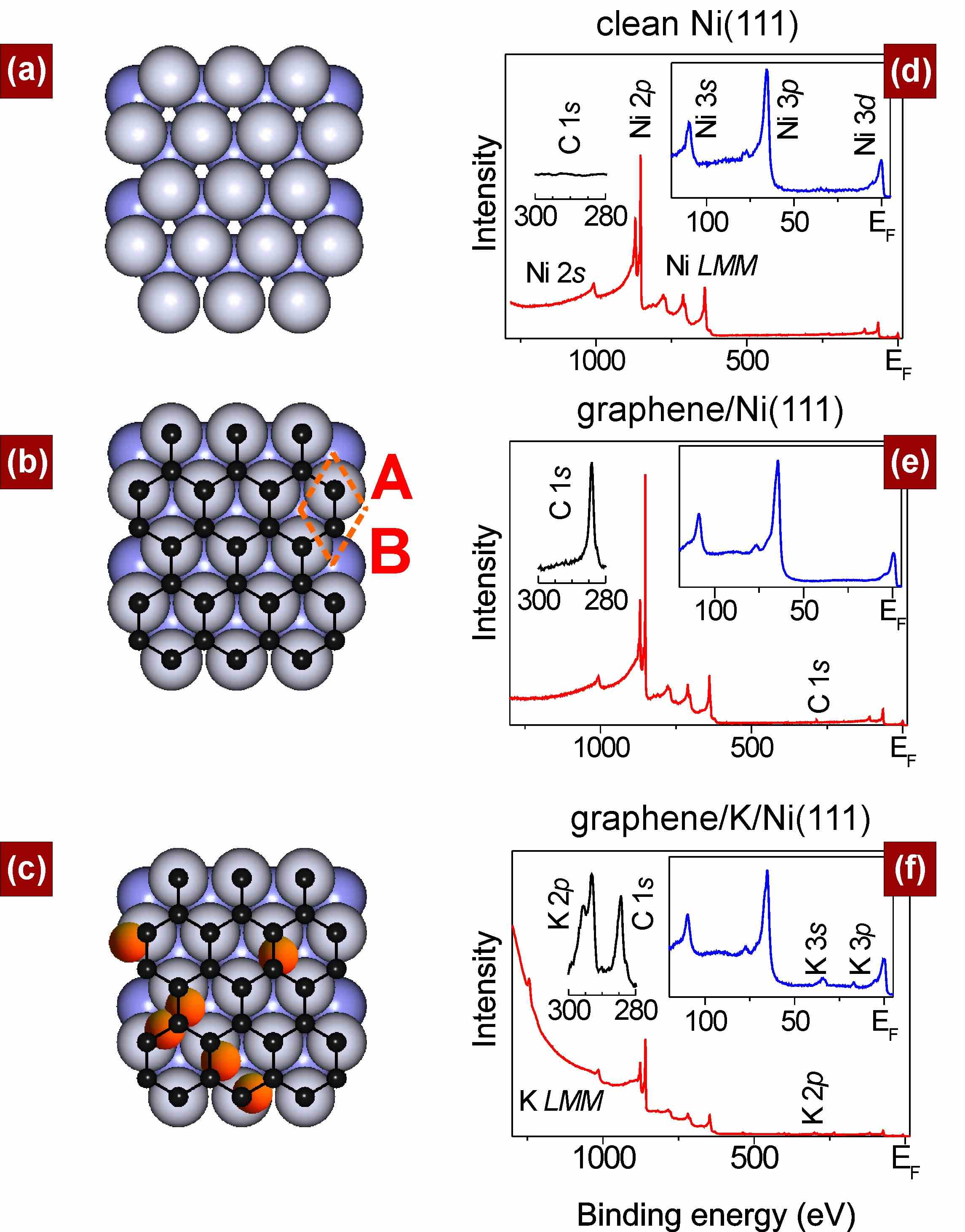}
\end{tabular}
\caption{ Epitaxial growth and functionalization of a graphene
monolayer: (a) Clean Ni(111) surface. (b) Graphene monolayer grown
by CVD. The unit cell with the non--equivalent $A$ and $B$ atoms
is indicated. (c) Potassium atoms (orange) intercalate between the
Ni(111) and the graphene. The corresponding XPS spectra for (d) a
clean Ni(111) surface, (e) an epitaxially grown graphene monolayer
on Ni(111) and (f) a Potassium intercalated graphene monolayer
with K/C=0.69.} \label{fig:leed}
\end{figure}
Fig.~\ref{fig:leed}(a)--(c) show the sample preparation
schematically. In the beginning the electronic and crystalline
structures of the clean Ni(111) surface are inspected.
Subsequently, we perform chemical vapour deposition (CVD) of
propylene in order to grow the graphene layer catalyzed by
Ni(111)~\cite{oshima94-graphene,molo01-graphene}. After that we
evaporate potassium onto the sample. Fig.~\ref{fig:leed}(d)--(f)
displays XPS spectra taken at each stage of the experiment.
Inspecting these spectra briefly, it is obvious that appearance of
the graphene layer is accompanied by a sharp peak at $\sim$ 285 eV
BE due to electron emission from the $1s$ core-level of carbon.

\begin{figure}[t]
\begin{tabular}{c}
\includegraphics[width=9cm]{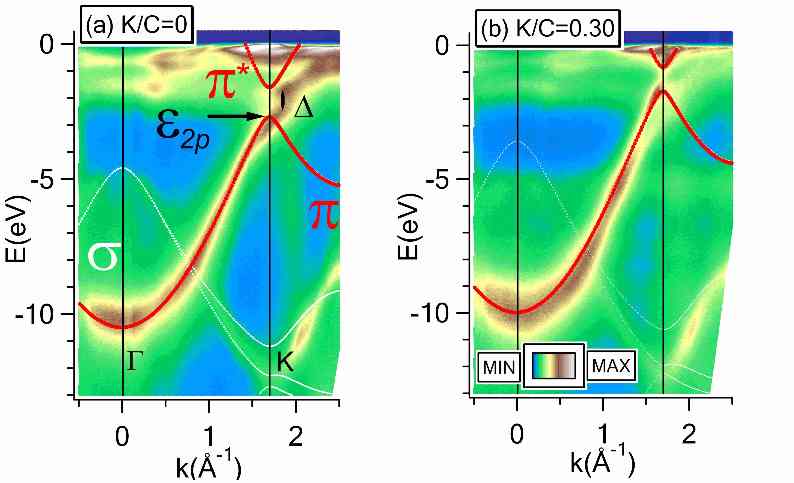}
\end{tabular}
\caption{ARPES intensities of the band structure of (a) a pristine
graphene monolayer and (b) a doped graphene monolayer. The red and
white lines denote a TB calculation for $\pi$ and $\sigma$ bands
(for parameters see Table~\ref{tab:tb}). The on--site energy
$\epsilon_{2p}$ and the gap $\Delta$ are depicted.}
\label{fig:arpes1}
\end{figure}

In Fig.~\ref{fig:arpes1} the electronic structure of pristine and
potassium doped graphene/Ni(111) is shown. Inspecting the topology
of the band dispersions one can make the following general
observations: (i) for K/C=0 the $\pi$ band structure is shifted
down in binding energy by $\sim$2.7~eV with respect to what is
expected from theory for an isolated graphene monolayer; (ii) for
K/C=0.30 the intensity from Ni $3d$ states with a flat dispersion
close to the Fermi level is suppressed and the $\pi$ and $\pi^*$
bands at $K$ point move up in BE. Thus for the K/C=0.30 the
observed graphene electronic bands are that of a doped graphene
layer with a rather small interaction with the Ni substrate. The
first point is explained by a strong covalent bonding between the
graphene layer and the Nickel surface. The second point indicates
that intercalated potassium atoms gradually extends the interlayer
distance without affecting the morphology of the graphene
monolayer. This decreases the overlap between wave functions of C
$2p_z$ and Ni $3d_{z^2-r^2}$ states oriented perpendicular to the
surface, decreasing the strong interlayer interaction of graphene
and the substrate. The proof for intercalation of potassium to the
Nickel/graphene interface comes from the intensity ratio of the
photoemission from $\pi$ states of graphene to $d$ states of
Ni(111). We observe a strong increase of this intensity ratio.
This clearly indicates penetration of potassium atoms to the
graphene/nickel interface. Since we do not observe an intercalant
band from the potassium 4$s$ electrons we conclude that we have
complete charge transfer of one electron per potassium atom. The
amount of charge transfer is strongly dependent on the carbon
material and the intercalant atom and for different systems one
might expect a different degree of charge
transfer~\cite{rukola08-kc8lda}.

\begin{figure}[b]
\begin{tabular}{c}
\includegraphics[width=8cm]{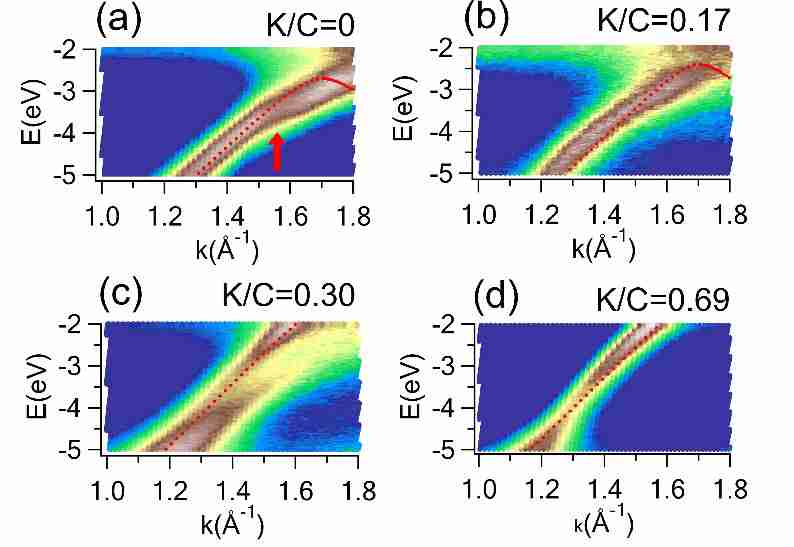}
\end{tabular}
\caption{The region of the hybridization kink in the $\pi$ band of
graphene for different doping levels as indicated by the K/C
ratio. The kink is denoted by a red arrow in (a). Red dots depict
the tight--binding calculations.} \label{fig:arpes2}
\end{figure}

In order to describe the electronic structure of the intercalated
graphene monolayer we perform TB calculations of the $\pi$ and
$\sigma$ bands taking into account nearest neighbor
interactions~\cite{s617}. In order to describe the influence of
the Ni(111) surface and intercalated K ions on the electronic
structure of the graphene monolayer we adjusted the values of the
on--site C $2p$ energies for the $A$ and $B$ atoms of the graphene
unit cell [see Fig.~\ref{fig:leed}(b)], i.e. we employed the rigid
band shift model. The $\pi$ band Hamilton and overlap matrices are
given by
\begin{equation}
\label{eq:hs}
\begin{array}{ll}
H({\bf k}) =\left (
\begin{tabular}{cc}
$\epsilon_{2p}$ & $\gamma_0f({\bf k})$ \\
$\gamma_0f^*({\bf k})$& $\epsilon_{2p}+\Delta$
\end{tabular}
\right ) \end{array}
\end{equation}
and
\begin{equation}
\label{eq:hs}
\begin{array}{ll}
S({\bf k}) =\left (
\begin{tabular}{cc}
1 & $sf({\bf k})$ \\
$sf^*({\bf k})$& $1$
\end{tabular}
\right ), \end{array}
\end{equation}
respectively. Here we use
\begin{equation}
f({\bf
k})=\exp(i\frac{k_xa}{\sqd})+2\exp(-\frac{ik_xa}{2\sqd})\cos(\frac{k_ya}{2}).
\label{eq:fp1p2p3}
\end{equation}
\begingroup
\squeezetable
\begin{table}
\begin{tabular}{c|c|c|c|c}
\hline \hline K/C ratio&0&0.17&0.3&0.69\\
\hline
$\epsilon_{2p}$& -2.7& -2.4& -1.6 & -1.5\\
$\Delta$ & 0.9 & 0.8 & 0.9 & 0.8\\
$\gamma_0$&3.2&3.2&3.4&3.3\\
\hline \hline
\end{tabular}
\caption{Parameters for the on--site energy ($\epsilon_{2p}$), gap
($\Delta$) and nearest neighbour in--plane overlap of C~$2p_z$
orbitals ($\gamma_0$) of the TB calculation shown in
Fig.~\ref{fig:arpes1} and Fig.\ref{fig:arpes2}. All values are
given in eV.}\label{tab:tb}
\end{table}
\endgroup
At the $K$ point of the two--dimensional Brillouin zone of
graphene we obtain $f({\bf K})=0$ and thus the Eigenvalues at $K$
are given by the diagonal elements of $H({\bf k})$, i.e.
$\epsilon_{2p}$ and $\epsilon_{2p}+\Delta$, respectively. The
electronic band structure is given by solving the Schr\"odinger
equation $H({\bf k}){\bf c}({\bf k})=E({\bf k})S({\bf k}){\bf
c}({\bf k})$, where $E({\bf k})$ and ${\bf c}({\bf k})$ denote the
Eigenvalues and wave function coefficients, respectively. We
managed to reproduce the topology of the graphene electronic
states by fitting $\epsilon_{2p}$, $\Delta$ and the $\pi$ overlap
($\gamma_0$) to the experimental data. Other parameters related to
the $\sigma$ bands such as the $2s$ on--site energy and $sp$
overlap need not to be changed since the hybridization of $\sigma$
states of graphene with \textit{d}-states of the Ni(111) surface
is rather weak. The values of the parameters used in the TB
calculation are summarized in Table~\ref{tab:tb}. Our calculations
demonstrate that with proceeding potassium intercalation
$\epsilon_{2p}$ decreases distinctly by 1.2~eV in direction of
E$_F$. At the last stage of the experiment, i.e. for complete
intercalation, $\epsilon_{2p}$ was found to be at 1.5~eV below
E$_F$. The values for $\Delta$ and $\gamma_0$ do not reveal
significant changes with increasing K/C ratio. Now, let us take a
closer look at the shape of the $\pi$ band between $\Gamma$ and
$K$ points shown in Fig.~\ref{fig:arpes2}. Obviously, in the
vicinity of the $\Gamma$ point the dispersion of the
experimentally derived $\pi$ band is accurately reproduced by the
TB calculation. However, close to the $K$ point this agreement is
strongly affected by the value of the K/C ratio. For the case of
the pristine graphene/Ni(111) structure (K/C=0) the "kink" shape
of this band is clearly visible. The BE of the "kink" is too high,
about 3.5~eV, to attribute it to electron-plasmon or
electron-phonon interactions. Accurate consideration of this
picture suggests that the $\pi$ band is pushed downwards by the Ni
$3d$ bands. Thus one can suppose that the "kink" is a product of
hybridization between graphene and nickel states.

Details of the dispersion of $\pi$ band close to the point where
the "kink" was monitored for K/C=0 are shown in
Fig.~\ref{fig:arpes2} as a function of the K/C ratio. Note that
with increasing amount of intercalated potassium atoms the
strength of the "kink" is gradually reduced. For the case of
K/C=0.3 one can distinguish a second component of the $\pi$ band.
This fact may be interpreted as follows: there are certain sample
positions where potassium intercalates already underneath the
graphene, reducing its interaction with the Ni(111) surface. Thus,
at this stage both phases, a strongly hybridized and an
intercalated graphene layer with small substrate interaction,
coexist. For the case K/C=0.69 the "kink" disappears and the usual
steep slope behavior of the $\pi$ band appears. Further increasing
K/C does not reveal significant changes.

We discuss now the present experimental and theoretical results.
The derived fit parameters for the $\pi$ bands imply the following
physical interpretation: by intercalation of potassium atoms to
the graphene/Ni(111) interface, the strong hybridization between C
$2p_z$ and Ni $3d_{z^2-r^2}$ orbitals is weakened. Furthermore,
when the graphene layer is lifted up, it is no longer attached to
the Ni(111) surface and the lattice constant of graphene might
relax to a slightly lower value. These two effects are expected to
decrease $\epsilon_{2p}$ and increase $\gamma_0$. While the
increase of $\epsilon_{2p}$ is clearly observable, the increase of
$\gamma_0$ is in the range of the experimental error. The reason
for this is probably that the difference in lattice constants is
too small (1.42$\rm\AA$ for C-C and 1.44$\rm\AA$ for Ni-Ni) to
produce a noticeable change in $\gamma_0$. The fact that $\Delta$
does not change over the whole doping range is quite different to
the case of bilayer graphene. In the case of bilayer graphene on
SiC the gap is equal to the doping dependent difference in the
on--site energy for the two
layers~\cite{rotenberg06-graphite_bilayer}. In the case of a
monolayer, the gap is given by the difference of the on--site
energy for the $A$ and $B$ atom in the graphene unit cell. Since
$A$ and $B$ atoms occupy non--equivalent sites on the Ni(111)
surface [see Fig.~\ref{fig:leed}(b)] this can be easily understood
for the case of K/C=0. For higher doping, the absence of a change
in the gap might thus be related to a superstructure of the
potassium atoms on the Ni(111) surface.

The $\pi$ valence bandwidth is given by $3\gamma_0$ and the
experimentally derived values for $\gamma_0$ in this work are
about 20\% larger than the LDA derived
ones~\cite{alex06-correlation}. We thus conclude that the increase
in the experimental bandwidth is due to electron--electron
correlations, similar to the case of
graphite~\cite{alex06-correlation}. The valence bandwidth of
graphene is not affected by the fact that the graphene monolayer
is hybridized with the metal surface since it is identical with
the graphite bandwidth. Such a result is surprising since one
might expect that the graphene/metal interface enhances the
screening of carriers in graphene and thus leads to a better
agreement with the LDA. Hopefully, these new findings will
stimulate further theoretical considerations.

Concerning other spectroscopic techniques it would be interesting
to probe the electronic structure of pristine and functionalized
graphene monolayers by optical spectroscopy such as resonance
Raman. The (double) resonance Raman process involves
$\pi\rightarrow\pi^*$ transitions~\cite{c887} and in order to
fulfill the resonance condition it is clear that the laser energy
h$\nu$ must be larger than $2|\epsilon_{2p}|-\Delta$. From ARPES
experiment we can predict the resonance condition which yields
h$\nu$=4.5 eV for K/C=0 and h$\nu$=2.2 eV for K/C=1.02 as a lower
limit for the laser energy in resonance Raman measurements.

Finally, we would like to add one more point. It is reasonable to
anticipate that alkali-metal intercalated high-quality graphene
layers on Ni(111) surfaces bring up the opportunity to use them
for the preparation of top quality graphene layers on different
substrates. Since the intercalation liberates the graphene from
strong covalent bonding with Ni(111), it would be feasible to peel
off intercalated graphene from the substrate.

\textit{In summary}, we have investigated modifications of the
electronic structure of the graphene layer on the Ni(111)
substrate upon gradual intercalation of potassium metal. We found
that the hybridization strength between graphene and Ni states can
be monitored successfully by the "hybridization" kink of the $\pi$
band distribution while the charge transfer follows a rigid band
shift model in perfect agreement with results of TB calculations.
We anticipate that the graphene/Ni(111) structure could be used
successfully as a model system capable to provide valuable insight
into the mechanisms of electron correlations and many-body
interactions in solids.

A.G. acknowledges a Marie Curie Individual Fellowship (COMTRANS)
from the European Union. D.V. acknowledges the Deutsche
Forschungsgemeinschaft (SFB 463) for projects TP B4 and TP B16.
Fruitful discussions with Thomas Pichler, Martin Knupfer, Clemens
Laubschat and Serguei Molodtsov are gratefully acknowledged.

\end{document}